\documentclass[twocolumn,showpacs,prd,floatfix]{revtex4}

\usepackage{graphicx}
\usepackage{dcolumn}
\usepackage{bm}
\usepackage{amsmath}
\usepackage{epsf}

\begin{document}

\newcommand{\beq}{\begin{equation}}
\newcommand{\eeq}{\end{equation}}
\newcommand{\bea}{\begin{eqnarray}}
\newcommand{\eea}{\end{eqnarray}}
\newcommand{\lsim}{\stackrel{<}{\scriptstyle \sim}}
\newcommand{\gsim}{\stackrel{>}{\scriptstyle \sim}}
\def\bone{$B^{(1)}$}
\def\bone{B^{(1)}}
\def\etal{{\it et al.~}}
\def\eg{{\it e.g.~}}
\def\ie{{\it i.e.~}}
\def\DM{dark matter~}
\def\DE{dark energy~} 
\def\GC{Galactic center~} 
\def\susy{SUSY~}

\title{Effects of atmospheric electric fields on detection of
  ultrahigh-energy cosmic rays} 

\author{Alexander Kusenko and Dmitry Semikoz
}
\affiliation{Department of Physics and Astronomy,
UCLA, Los Angeles, CA 90095-1547, USA}

\vspace{0.5truecm}
\begin{abstract}
We show that atmospheric electric fields may affect the cosmic ray
observations in several ways and may lead to an overestimation of the
cosmic ray energies.  The electric field in thunderclouds can be as high as
a few kV/cm.  This field can accelerate the shower electrons and can feed
some additional energy into the shower.  Therefore, ground array
observations in certain weather conditions may overestimate the energy of
ultrahigh-energy cosmic rays if they don't take this effect into account.
In addition, the electric field can bend the muon trajectories and affect
the direction and energy reconstruction of inclined showers.  Finally,
there is a possibility of an avalanche multiplication of the shower
electrons due to a runaway breakdown, which may lead to a significant
miscalculation of the cosmic ray energy.

\end{abstract}

\pacs{96.40.-z, 95.55.Vj, 95.85.Ry, 98.70.Sa  
\hfill 
UCLA/04/TEP/44}
\maketitle

\vspace{1truecm}

\section{Introduction}

Ultrahigh-energy cosmic rays (UHECR) produce extensive air showers in the
atmosphere, which are observed by two different techniques: (1) optical
detection of fluorescent light, and (2) surface array detection of charged
particles in the shower.  The first technique, employed, for example, by
HiRes detector \cite{hires}, can be used exclusively in good weather, while
the second technique, used, for example, by the AGASA
experiment\cite{AGASA}, is usually believed to afford accurate observations
regardless of the atmospheric conditions.  The new Pierre Auger experiment
is using a combination of the two techniques \cite{AUGER}.

On a clear day the atmosphere is permeated by the electric fields of order
a few volts per cm~\cite{intro}.  However, in thunderclouds the electric
field reaches much higher magnitudes, up to a few kV/cm~\cite{el_book}.
Balloon measurements~\cite{Marshall} show the electric fields ${\cal E}>
1$~kV/cm, at altitudes 0-12 km, which may switch polarity at several
altitudes.  A shower developing in the atmosphere and going through a
thundercloud may, therefore, pass between layers with either polarity at
some angle.  The electric field along the path of a random shower may have
a magnitude of a few kV/cm and an arbitrary direction, or alternating
direction.  As we discuss below, the electron energy gain in the electric
field of a few kV/cm can be comparable to the energy losses due to
ionization.  It, therefore, important to examine what effect this may have
on detection of UHECR.

Several experimental studies confirm the influence of electric fields on
secondary particles in a shower.  First, observations of low-energy cosmic
ray electrons show short-term variations during thunderstorms
\cite{thunder_low_energy}.  A similar effect has been observed for muons
with a higher energy,
$E>100$~MeV~\cite{thunder_low_energy,thunder_low_energy2}, but it is not
nearly as strong as the effect on the low-energy electrons.  Second, it has
been established that giant electron-gamma bursts are triggered in
thunderclouds by the passage of an extensive air shower (EAS) from a
$10^{16}$ cosmic ray \cite{electr_thunder}.  EAS were measured at the Tien
Shan Mountain in coincidence with detection of a radio signal from the giant
electron-gamma bursts. Theory of these bursts, based on the phenomenon of
runaway breakdown, has been developed in recent years~\cite{uspekhi}.  It
has was also suggested that 
acceleration of secondary electrons from EAS in thunderclouds can be used
for radio detection of cosmic rays with energies
$E>10^{17-19}$~eV~\cite{radio_high_energy}.

We will show that the atmospheric electricity associated with thunderclouds
may affect the ground array observations of $E>10^{19-20}$~eV cosmic rays
in several ways.  First, acceleration of the shower electrons can alter the
low-energy spectrum and the number of particles in EAS at ground level.
This could lead to errors in energy determination by a ground array.
Second, the muon trajectories may be deflected by the atmospheric electric
fields, which could lead to errors in reconstruction of energy and
direction of inclined showers. Finally, a runaway breakdown and
discharge triggered by an UHECR EAS may produce a dramatic increase in the
number of electrons detected by the surface array.

\section{Energy of the electromagnetic component}

\begin{figure}[ht]
\centerline{\epsfxsize=3.5in\epsfbox{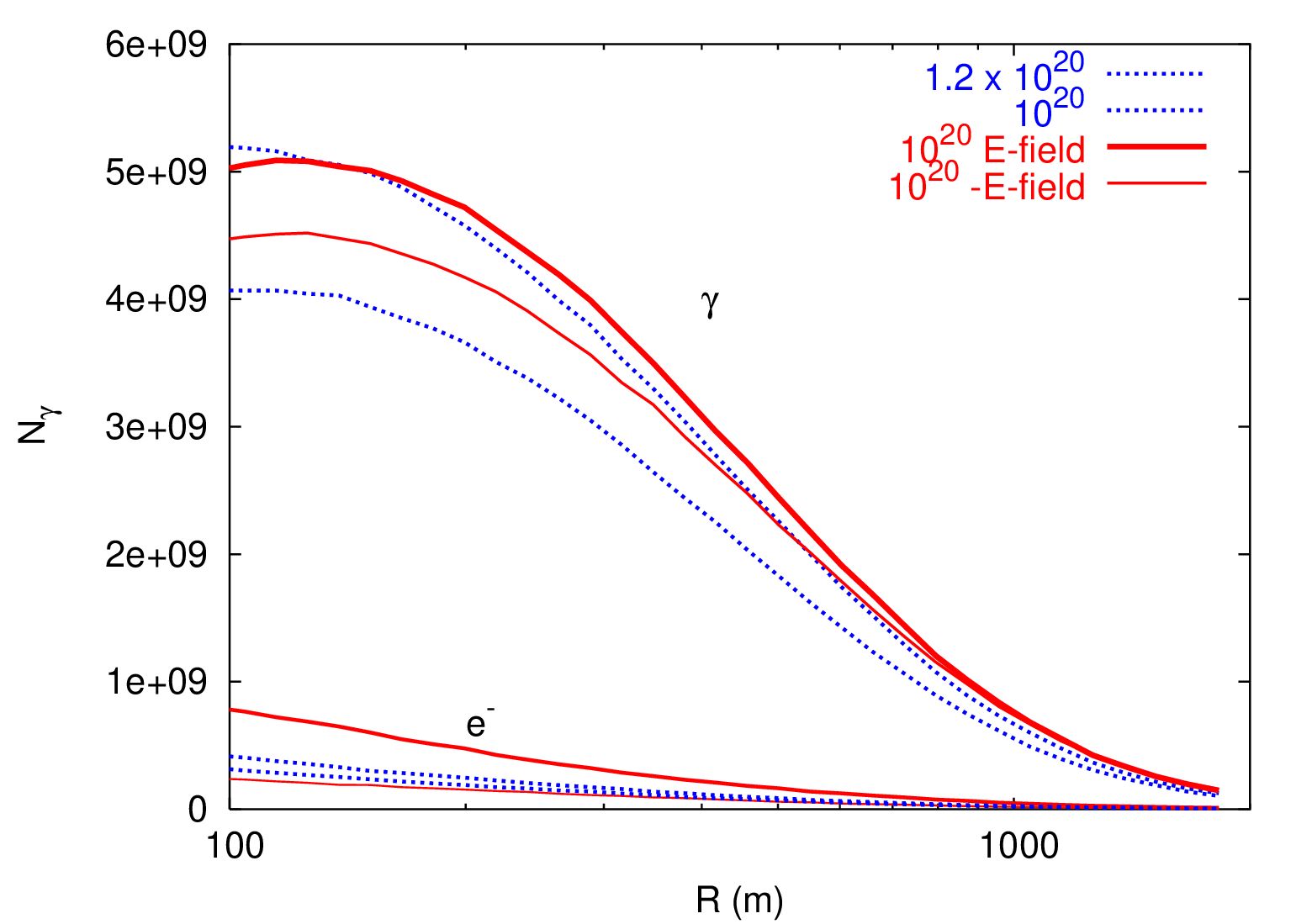}}   
\centerline{\epsfxsize=3.5in\epsfbox{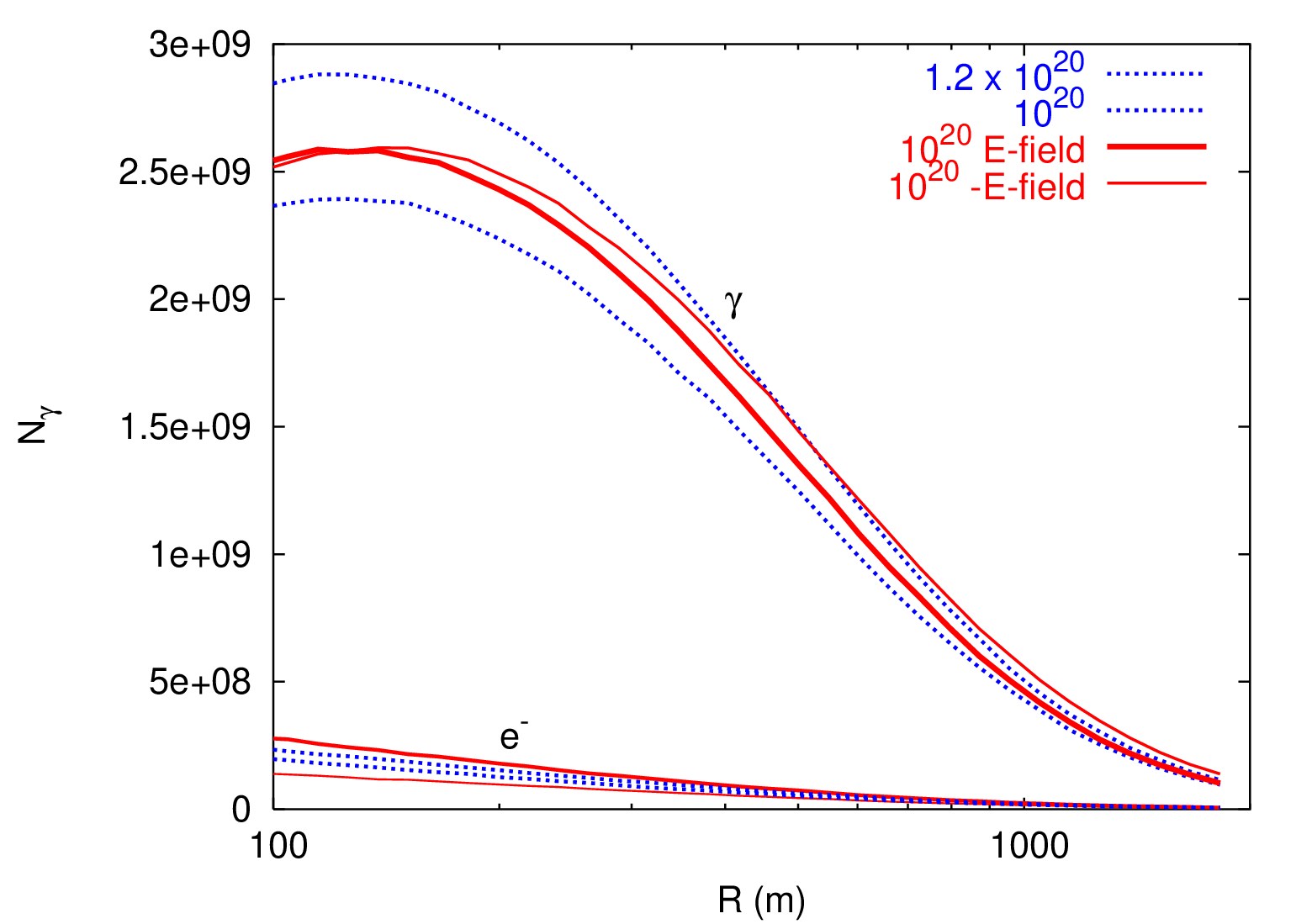}}   
\caption{ Effect of a uniform atmospheric electric field on the shower
profile for a proton primary (upper plot) and an iron primary (lower plot).
The dotted lines show AIRES simulations of the $10^{20}$~eV
(low line) and $1.2\times10^{20}$~eV (upper line) showers, respectively, in
the absence of atmospheric electric field. The thick and thin solid lines
correspond to a $10^{20}$~eV shower developing in the presence of a uniform
electric field $\pm 1$~kV/cm.  The electric field makes the $10^{20}$~eV
shower look more like a $(1.1 - 1.2)\times10^{20}$~eV shower.  }
\label{figure:R}
\end{figure}

\begin{figure}[ht]
\centerline{\epsfxsize=3.5in\epsfbox{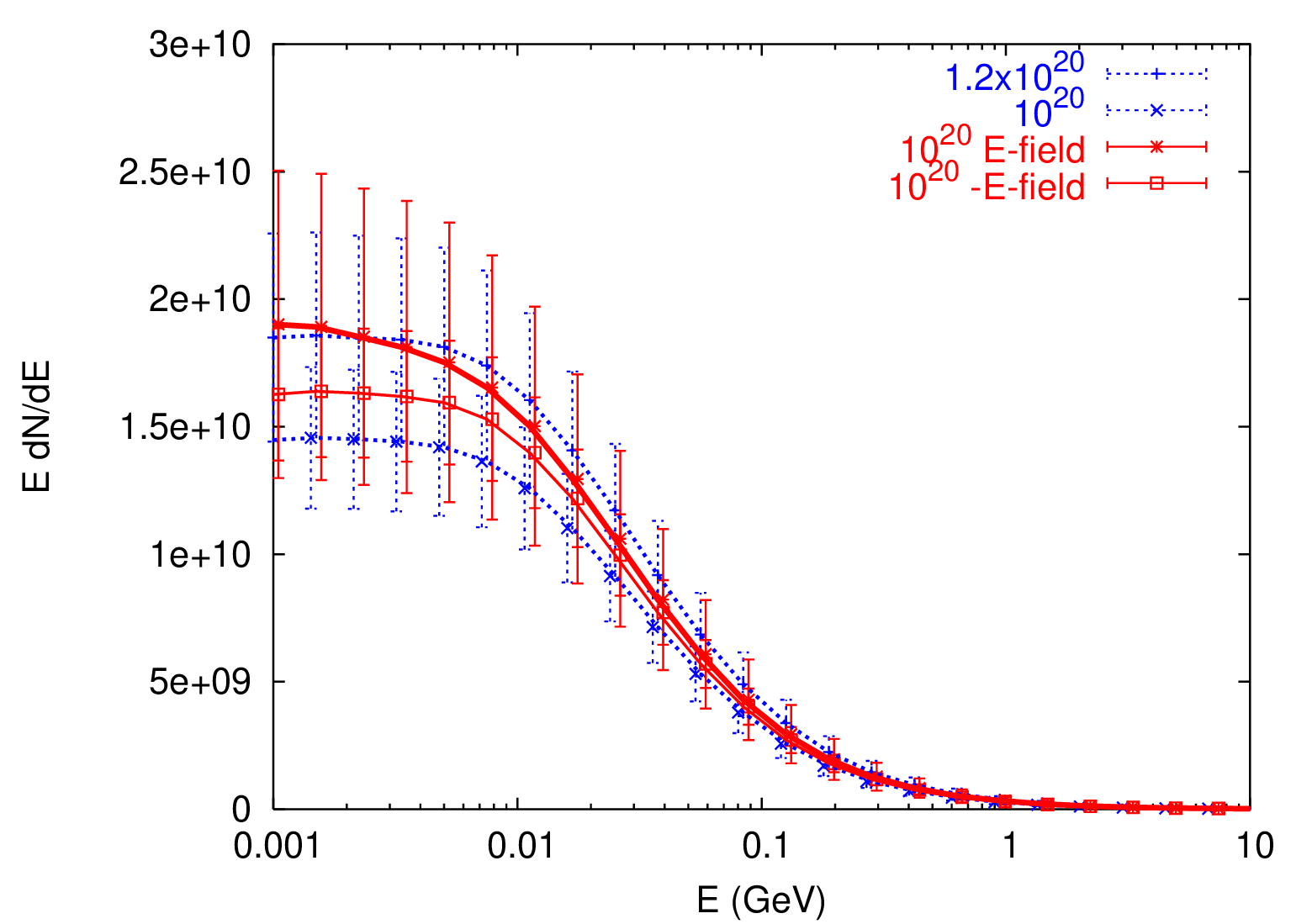}}   
\centerline{\epsfxsize=3.5in\epsfbox{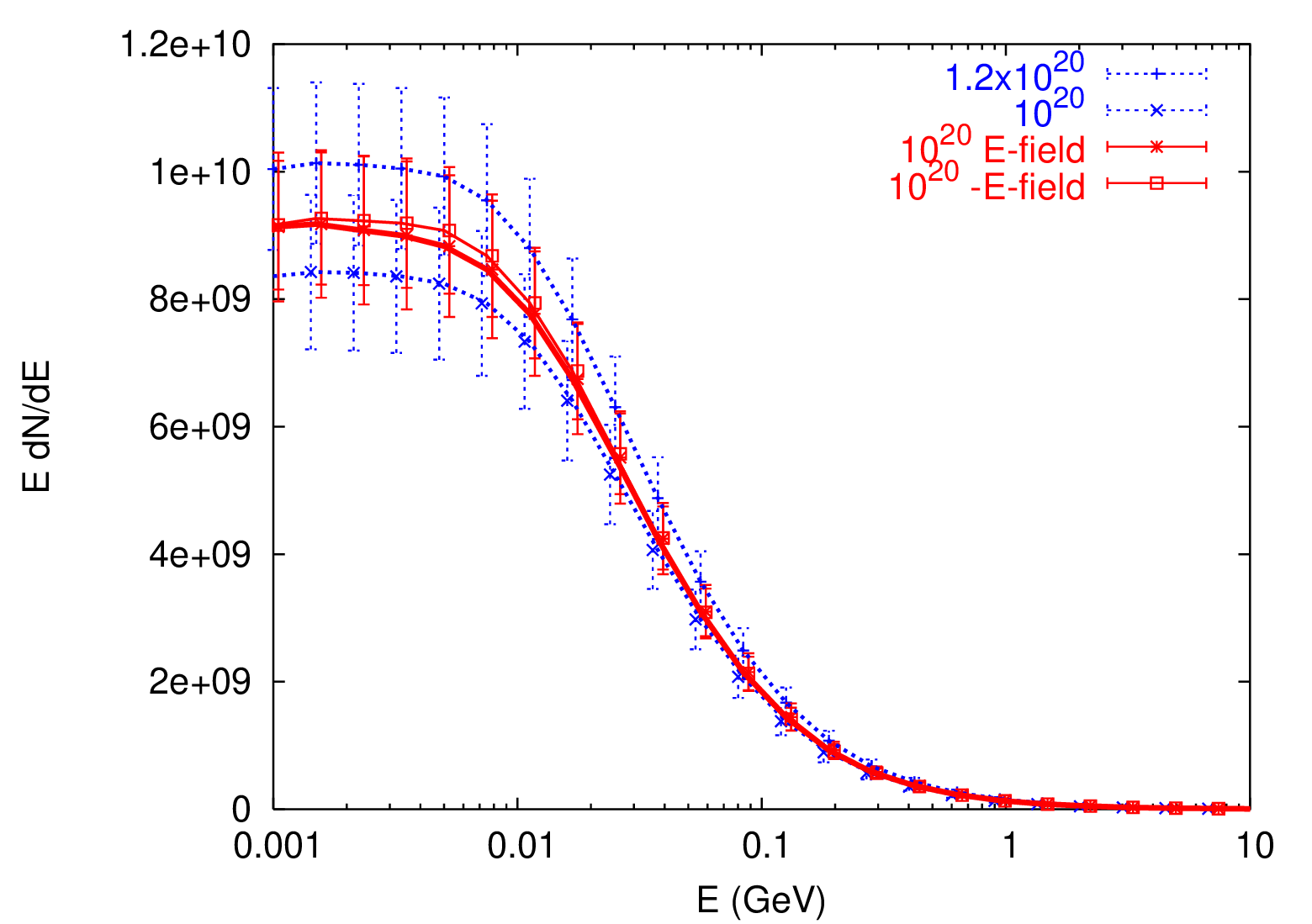}}   
\caption{ Effect of atmospheric electric field on the photon and electron
spectra.  The the dotted lines represent the spectra of $10^{20}$~eV and
$1.2\times10^{20}$~eV showers, respectively, in the absence of atmospheric
electric field.  The thick and thin solid lines correspond to a
$10^{20}$~eV shower developing in the presence of a uniform electric field
$\pm 1$~kV/cm.  In the presence of electric field, the apparent energy of
the shower is higher than it's actual energy. 
}
\label{figure:E}
\end{figure}

An electron or positron with momentum $\vec{p}=\vec{e}_x p_x + \vec{e}_y
p_y$ moving in the atmospheric field ${\cal E}(x,y)$ is described by the
following system of equations:
\beq
\left \{
\begin{array}{l}
\dot{p_x}=-\kappa(y) \frac{p_x}{p} \\
\dot{p_y}=e{\cal E}(x,y) -\kappa(y) \frac{p_y}{p}, 
\end{array}
\right.
\label{eqn_motion}
\eeq
where $\kappa$ accounts for the ionization losses.  Here we have assumed
$\kappa $ to be independent of energy (a good approximation for the shower
electrons).  Of course, $\kappa$ depends on the air density, and, hence, on
the height $y$ in the atmosphere.  The electric field is also a function of
coordinates.  

For a vertical shower in a vertical electric field ${\cal E}={\cal E}(y)$,
the horizontal 
component of the momentum vanishes, $p_x=0$, and the effect of the electric
field amounts to reducing the energy losses of the shower in the
atmosphere.  The equation of motion is

\beq
\dot{p_y} \approx \frac{d}{dy} p_y=e{\cal E}(y)-\kappa(y), 
\eeq
where we have taken, as usual $c=1$. The solution of this equation, 
\beq
p_y(y)= p_{y,0} - \int_{y_0}^{y} (e{\cal E}(\xi)-\kappa(\xi)) d\xi, 
\label{integral} 
\eeq 
depends on the initial momentum of an electron and on the form of $(e{\cal
E}(y)-\kappa(y))$, where $\kappa(y)$ is a function of density only. From
the form of this solution it is clear that, for a vertical shower, the
electric field can reduce the energy losses and to make the atmosphere
effectively thinner.  Of course, if the electric field is larger than
$\kappa/e$, the electron gains energy faster than it loses energy.  This
possibility will be discussed below.

The energy losses of the shower electrons due to ionization are 
\beq
\kappa = \frac{dE}{dy}\approx \left ( \frac{\rho}{0.0013 {\ \rm g\,
    cm}^{-3}} \right 
) \left [\ln 
  \frac{E}{m_e} 
    +6.3 \right ] \ \frac{\rm keV}{\rm cm} . 
\eeq
This should be compared with the atmospheric electric field, which reaches
the values 
\beq
{\cal E }\gsim 
1 \, \frac{\rm kV}{\rm cm} 
\eeq
in thunderclouds~\cite{el_book}.  Obviously, the effect of the electric field
is non-negligible.

The experimental determination of UHECR energy is based on the
reconstruction of the total energy of photons and electrons at the shower
maximum based on the observed total energy at ground level.  The
extrapolation to shower maximum involves, effectively, solving the
evolution of shower energy due to the energy losses as well as the change
in the spectrum and the number of particles due to a number of elementary
processes.  Most of the particles reaching the ground are photons; in
particular, there are more photons than electrons at ground level by an
order of magnitude.  These photons originate from (i) bremsstrahlung of
electrons, (ii) annihilations of positrons.  If the electrons are
accelerated by the field, the positrons are slowed down, and vise versa.
The electric field along the path of a shower can have either polarity.
The electric field may also have different directions at different
heights~\cite{el_book}.  On average, there is more positive charge in the
higher layers of the atmosphere, and the ionosphere is almost always
positively charged.  However, the thunderclouds is where the field reaches
a high magnitude.  The electric field in a cloud is not necessarily
uniform.  It can have arbitrary direction along the shower path. 

If the electrons are accelerated, the atmosphere is effectively thinner for
them.  Therefore, these electrons can produce the bremsstrahlung photons
over a longer period of time, and the resulting photons have, on average, a
higher energy.  If positrons are accelerated, they have a higher average
energy when they annihilate, and, in addition, they annihilate closer to
the ground, on average.  These are independent and, perhaps, competing
effects, so {\em a priori} it is difficult to know which of them dominates.
In addition, one has to keep in mind that the ground detectors are
sensitive to particles about 0.6-1.2~km away from the shower axis.  It is
difficult to evaluate the overall effect of the electric fields
analytically.  We have employed a numerical calculation using
AIRES~\cite{AIRES} software package to simulate the effects of the
atmospheric electric field.  To this end we changed the electron and
positron energy loss rate in accordance with eq. (\ref{integral}), assuming
a constant electric field along the direction of the shower. It turns out
that both the acceleration of electrons and positrons give similar effect.

We illustrate the effect in Figures~\ref{figure:R} and \ref{figure:E}.  In
Fig.~\ref{figure:R} we show the number of photons and electrons reaching
the ground as a function of distance from the shower core.  The upper plot
corresponds to a proton primary, while lower plot is for iron.  Dotted
(blue) lines present shower profiles with a primary energy $10^{20}$~eV
(lower line) and $1.2\times10^{20}$~eV (upper line). Thick (thin) solid
line corresponds to the case of a $10^{20}$~eV shower affected by a
1~keV/cm electric field in the direction for which the electrons
(positrons) are accelerated.  One can see that, in the presence of the
electric field, the number of photons increases, and the $10^{20}$~eV
shower begins to resemble a $1.2\times10^{20}$~eV shower.  Although it
makes a difference whether electrons or positrons are accelerated
(decelerated), the overall effect increases the apparent shower energy in
both cases.  One can also see from Fig.~\ref{figure:R} that number of
electrons and positrons reaching the ground changes with the direction of
the field, but that their number in any case is negligible in comparison
with the number of photons. 

In Fig.~\ref{figure:E} we show the energy distribution of photons and
electrons in the shower at ground level.  The curve markings are similar to
those in Fig.~\ref{figure:R}.  In the presence of an electric field, the
apparent shower energy may seem higher than the actual energy by as much as
20\%.

We emphasize that our simulations were very crude: we assumed a uniform
electric field along the shower path. In reality, the electric field
configuration can be much more complicated, and this can either increase or
decrease the significance of the effect.  Also, in these numerical
simulations we neglected the effect of an electric field on the muon
trajectories.  More detailed simulations are required for more reliable
results.  It would be desirable to measure the electric fields above the
experimental site and use these measurements in the data analysis.
Measuring the electric field at ground level is relatively straightforward.

\section{Muon arrival directions} 

The Pierre Auger detector is sensitive to both electrons and muons in the
shower.  For a highly inclined shower, the electromagnetic component dies
out before the shower reaches the ground, and, therefore, the data analysis
is based on muons alone.  The arrival times of muons are used to
reconstruct the angle of the shower.  Here the effects of the atmospheric
electric field must also be taken into account because the electric field
bends the muon trajectories.

Let us approximate $\kappa $ by a constant (constant density atmosphere),
and let us consider a uniform electric field. For a small 
field, the angle is approximately constant and is close to the initial
value $\theta_0$,
\beq
\tan \theta_0 =\frac{p_{x,0}} {p_{y,0}}.  
\eeq
As long as this is the case, the equations (\ref{eqn_motion}) have an
approximate solution
\beq
\left \{
\begin{array}{l}
{p_x}=p_{x,0}-\kappa \sin \theta_0 \, t  \\
{p_y}=p_{y,0} + e{\cal E} t -\kappa\cos \theta_0 \, t. \\
\end{array}
\right. 
\label{solu_muon}
\eeq
The components of the particle's velocity are 
\bea
\dot{x} =\frac{p_x}{\sqrt{m^2+p^2}} \\
\dot{y} =\frac{p_y}{\sqrt{m^2+p^2} } , 
\eea
where, as before, $p=\sqrt{p_x^2+p_y^2}$.  For the zenith angle one can write 
\beq
{\rm cotan}\, \theta =\frac{dy}{dx}=
\frac{\dot{y}}{\dot{x}}=\frac{p_y}{p_x}.  
\eeq
Using the approximate solution (\ref{solu_muon}), we get 
\bea
\dot \theta & = & \frac{d}{dt} \arctan \left ( \frac{p_x}{p_y} \right )   
\nonumber  \\ 
&  =& \frac{p_{x,0}(e {\cal E} - \cos \theta_0 \kappa)+ p_{y,0}\sin \theta_0 }{
p_0^2+2 A t +B t^2}, 
\eea
where 
$A=( e {\cal E} - \cos \theta_0 \kappa) p_{y,0}-\kappa \sin \theta_0
p_{x,0}  $ and $B= e^2E^2-2e E \kappa \cos \theta_0+\kappa^2 $.  Therefore,
the change in the angle is of the order of 
\beq
\Delta \theta \sim \frac{e {\cal E}t}{p} \sim 10^{-1} \left (\frac{3 {\rm
    GeV}}{p} \right ) 
\left (  
\frac{e E}{1 \, \frac{\rm kV}{\rm cm}}
\right )
\left (  
\frac{ct}{3 \, {\rm km}}
\right )
\eeq
for a uniform electric field $E$.  This change is non-negligible.  It 
introduces an error in the angle reconstruction of the arrival direction.
In addition, if one uses muons for energy estimation, this effect can
introduce a systematic error because one could miscalculate the length of
the shower path in the atmosphere.

\section{Runaway breakdown}

When the electric field in a thundercloud reaches some critical value of the
order a few keV, a runaway breakdown may generate a large number of
avalanche electrons with energies in the 0.1-10 MeV~range~\cite{uspekhi}.
The runaway electron avalanches are triggered by the seed electrons
in the extensive air showers initiated by the cosmic rays of energies
$10^{16}$~eV and higher~\cite{uspekhi}.  There is a mounting evidence that
this form of discharge is the reason why the electric field in thunderclouds
is limited to a few keV.  In the absence of cosmic rays the field could
reach values that are an order of magnitude greater than those observed in
clouds. 

If an electron avalanche accompanies an air shower observed by the ground
array, the number of electrons and photons registered by the detector could
increase significantly, and the cosmic ray energy can be greatly
overestimated.

The runaway breakdown requires the electric field to reach a critical
value~\cite{uspekhi}, which is  
\beq
{\cal E}_c \approx \frac{44 \, \pi e^3 Z n}{m_e} \approx 2.2\, \frac{\rm
  kV}{\rm   cm} 
\label{E_c}
\eeq
in the lower atmosphere. Here $n$ and $Z$ are the molecular density and the
electron number, respectively.  Since $10^{16}$~eV showers are apparently
sufficient to discharge the cloud, the field is unlikely to be critical at
the time and place of a (much less frequent) $10^{18}-10^{20}$~eV shower.
However, this probability is non-negligible and it is possible that some of
the high-energy showers can appear much more energetic to the ground array.
This effect should be taken into account by cosmic ray experiments.

\section{Conclusion}

We have shown that atmospheric electric fields may affect the energy
measurements by the ground arrays in certain weather conditions.  
Our simplified calculations show that, in the presence of thunderclouds,
one may overestimate the shower energy by as much as 20\%. 
In addition, the deflection of muons may affect the directional
reconstruction of inclined showers, which relies on muon arrival
times. Finally, if an UHECR shower is accompanied by a runaway breakdown,
the number of shower electrons  can increase dramatically, which
may lead to a significant miscalculation of the cosmic ray energy.

\section*{Acknowledgments} 

We thank K.~Arisaka, J.~Lee, T. Ohnuki, A.~Tripathi, and the rest of the
Pierre Auger group at UCLA for numerous discussions and comments.  We also
thank V.~Rubakov for helpful comments.  This work was supported in part by
the DOE grant DE-FG03-91ER40662 and the NASA ATP grant NAG~5-13399.

\end{document}